\documentclass[aps,prl,twocolumn,showpacs,superscriptaddress,floatfix,longbiblio
graphy]{revtex4-1}
\usepackage[utf8]{inputenc}
\usepackage{amssymb,amsmath}
\usepackage{graphicx}
\usepackage{braket}
\usepackage{siunitx}
\usepackage{xcolor}
\usepackage{physics}
\usepackage{hyperref}
\usepackage[normalem]{ulem}
\usepackage{orcidlink}

\AtBeginDocument{\RenewCommandCopy\qty\SI}

\hypersetup{
    colorlinks=true,
    linkcolor=blue,
    citecolor=blue,
    filecolor=blue,
    urlcolor=blue
}

\def\Na{^{87}\mathrm{Rb}}
\def\K40{^{40}\mathrm{K}}
\def\Na{^{23}\mathrm{Na}}
\def\kB{k_\mathrm{B}}

\def\spinup{\ket{\uparrow}}
\def\spindown{\ket{\downarrow}}

\def\ex{\mathbf{e}_x}
\def\ey{\mathbf{e}_y}
\def\ez{\mathbf{e}_z}

\begin{document}

\title{Prethermal ripplons in quenched binary Bose-Einstein condensates}
\author{Y. Geng\,\orcidlink{0009-0009-1937-409X}}
\affiliation{Joint Quantum Institute, National Institute of Standards and 
Technology, and University of Maryland, College Park, Maryland, 20742, USA}
\author{S. Eckel\,\orcidlink{0000-0002-8887-0320}}
\affiliation{National Institute of Standards and Technology, Gaithersburg, 
Maryland, 20899, USA}
\author{G. K. Campbell\,\orcidlink{0000-0003-2596-1919}}
\affiliation{Joint Quantum Institute, National Institute of Standards and 
Technology, and University of Maryland, College Park, Maryland, 20742, USA}
\author{I.~B.~Spielman\,\orcidlink{0000-0003-1421-8652}}
\email{ian.spielman@nist.gov}
\affiliation{Joint Quantum Institute, National Institute of Standards and 
Technology, and University of Maryland, College Park, Maryland, 20742, USA}
\date{\today}

\begin{abstract}
Prethermal states---long-lived quasi-equilibrium configurations---occur in a 
wide range of physical systems that exhibit fast dephasing or reconfiguration 
followed by slow relaxation towards thermal equilibrium.
We experimentally studied the 1D interface between immiscible 2D Bose-Einstein 
condensates (BECs); following a quench, the interfacial capillary waves 
(ripplons) quickly relaxed into a long-lived prethermal state characterized by 
a persistent non-equipartition of energy, evoking the classic 
Fermi-Pasta-Ulam-Tsingou problem.
We directly measured the interface's height profile as it evolved in time, 
identified the contribution of individual ripplon modes, and find that, though 
their individual amplitudes are thermally distributed, they are not in thermal 
equilibrium---high-momentum modes rapidly equilibrated with the bulk phonon 
modes of the 2D BEC, while lower momentum modes remained at elevated 
temperatures, forming a long-lived prethermal configuration for the whole 
system.
We attribute this to kinematic isolation: the absence of energy- and 
momentum-conserving relaxation processes in our system at ultracold temperature.
\end{abstract}

\maketitle

Quantum statistical mechanics is remarkably successful at predicting the 
equilibrium properties of many-body systems.
However, understanding how far-from-equilibrium systems approach equilibrium is 
a fundamental challenge in both quantum and classical physics, for all but the 
simplest of models.
Many systems undergo prethermalization~\cite{Mori2018}: quickly arriving at a 
quasisteady ``prethermal'' state that only slowly relaxes to global thermal 
equilibrium.
The Fermi-Pasta-Ulam-Tsingou (FPUT) problem is a key example: a chain of weakly 
nonlinearly coupled classical harmonic oscillators exhibits complex, 
nearly-integrable dynamics, and thermalization proceeds through high-order 
processes~\cite{Fermi1955,Benettin2013,Onorato2015}.
We observe prethermalization of ripplons at the 1D interface between immiscible 
2D Bose-Einstein condensates (BECs); their low-momentum modes form a long-lived 
prethermal state kinematically isolated by the absence of resonant energy- and 
momentum-conserving relaxation 
processes~\cite{Beliaev1957,Onorato2015,Zhang2021,Dereziski2024}, rather than 
by near-integrability.

Owing to prethermalization's simple underpinning, fast dephasing or 
reconfiguration followed by slow relaxation toward equilibrium, it is expected 
in settings ranging from inflationary cosmology and heavy-ion 
collisions~\cite{Berges2004, Kofman1994}, to classical FPUT-like systems and 
many-body quantum gases~\cite{Mori2018}.
Experiments on 1D ultracold 
gases~\cite{Kinoshita2006,Gring2012,Langen2015,Tang2018} and trapped-ion spin 
chains~\cite{Neyenhuis2017} observed long-lived prethermal states whose slow 
relaxation is tied to integrable or near-integrable dynamics.
Periodic driving provides another route, in which high-frequency Floquet 
systems prethermalize before eventual heating~\cite{RubioAbadal2020}.
In contrast to these examples, which are either isolated systems or protected 
by approximate conservation laws, our prethermal degrees of freedom are the 
ripplon modes of an interface embedded in a BEC; the bulk phonon modes 
therefore form a well-defined thermal bath that determines the interface's 
ultimate equilibrium state.

\begin{figure*}[t!]
    \centering
    \includegraphics{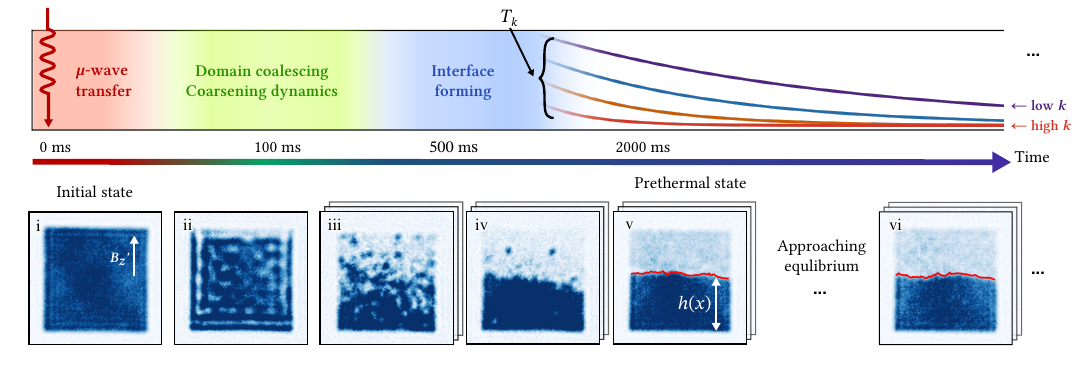}
	    \caption{
	    	The long road to equilibrium.
	        A schematic illustration of the relaxation process (top) is 
shown along with representative spin-selective images of the $\spinup$ density 
(bottom).
		        (i) A microwave pulse at $t=0$ transfers an initially 
spin-polarized BEC into an equal superposition of $\spinup$ (dark blue) and 
$\spindown$.
	        The orientation of a magnetic field gradient is indicated.
		        (ii) Initial spin structure forms nearly 
deterministically; (iii) small initial fluctuations then seed random droplet 
patterns that coalesce and coarsen.
		        (iv) After about $500\ {\rm ms}$ an interface forms and 
(v) becomes single-valued by $\num{2000}\ {\rm ms}$ (red curves).
	        Ripplons with different wavenumber $k$ combine to give the 
observed interface profiles; their occupation numbers $N_k$ (solid curves in 
the top panel) relax with drastically different timescales to their respective 
equilibrium values (dashed lines).
	    }
	\label{fig:schematics}
\end{figure*}

As shown in Fig.~\ref{fig:schematics}, our experiments began by using a 
microwave pulse to quench a spin-polarized BEC into an initially uniform 
far-from-equilibrium spin superposition, from which boundary-induced structure 
and amplified fluctuations seeded random droplet patterns that coalesced, 
coarsened, and finally formed two domains stabilized by a weak magnetic-field 
gradient~\cite{Huh2024,GengTao2025}.
The red curves in Fig. \ref{fig:schematics}(v) and (vi) show typical interfaces 
with height profiles $h(x)$. Height fluctuations result from excited interface 
modes, bosonic Bogoliubov quasiparticles called ripplons, with a range of 
wavenumbers $k$.
The amplitudes of the individual measured modes were consistent with thermal 
distributions, allowing us to assign an effective temperature $T_k$ to each 
ripplon.
As schematically shown in the top panel, high-$k$ modes rapidly thermalized 
with the bulk phonons, while low-$k$ modes remained out of equilibrium, with 
elevated mode temperatures, even for our longest hold times.
Our in-situ spin-selective imaging includes a faint, diffused signal above the 
interface [Fig.~\ref{fig:schematics}(v) and (vi)] resulting from a low density 
of thermally excited minority $\spinup$ atoms moving in the majority 
$\spindown$ BEC; their density can be directly related to the bulk temperature.

\vspace{3pt}\noindent
\textit{Experiment}---Our experiments began with optically trapped $\Na$ BECs 
with $\sim\num{1.6e6}$ atoms in the $\ket{F=1,m_F=-1}$ ground state in a 
quasi-2D square box potential in the $\ex\!-\!\ey$
plane, with side-length $L_0 = $ \qty{114}{\um}.
A blue-detuned Hermite-Gaussian beam provides strong confinement along $\ez$, 
yielding a transverse trap frequency
$\omega_z/2\pi =$ \qty{1.1}{\kHz}~\cite{GengTao2025,GengMurk2025}.
At $t = 0$, a microwave pulse transferred every atom into an equal 
superposition of the immiscible $\spindown \equiv \ket{F=1,m_F=-1}$ and 
$\spinup \equiv \ket{F=2,m_F=-2}$ states [Fig.~\ref{fig:schematics}(i)].
After a rapid process of domain formation and coarsening~\cite{Huh2024} 
[Fig.~\ref{fig:schematics}(ii)-(iii)], two spin domains formed on opposite 
sides of the box potential, stabilized by a small magnetic field gradient $B_z' 
= \qty{2.9(1)e-4}{\tesla\per\m}$ along
$\ey$; this yielded an interface nominally along $\ex$ [red curves in 
Fig.~\ref{fig:schematics}(v),(vi)].
The temperature of the condensate bulk was controlled by the final depth of the 
box potential $U_D$, which was slowly decreased while the interface formed.
After a hold time $t_{\rm hold}$, we selectively imaged the $\spinup$ density 
{\it in situ} by absorption imaging and extracted the interface height profile 
$h(x)$.

\begin{figure*}[t!]
    \centering
    \includegraphics{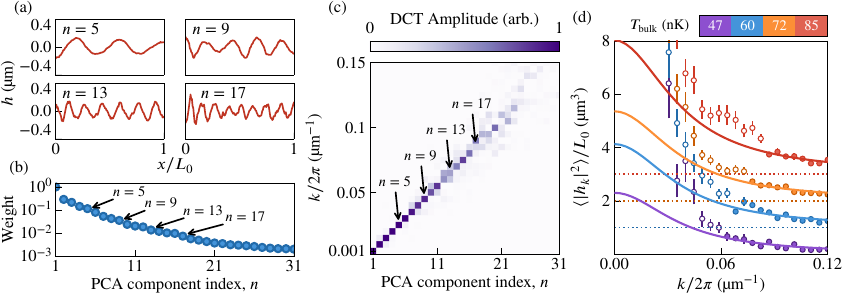}
    \caption{
        Fluctuations of a prethermal interface, with $t_{\rm hold} = 
\qty{2.0}{\second}$.
        (a) PCA components of $h(x)$ for $n= 5$, $9$, $13$, and $17$.
        (b) Normalized PCA weights.
        (c) DCT amplitudes of each PCA component.
        (d) Power spectrum $\langle |h_k|^2 \rangle$, measured at four bulk 
temperatures (vertically offset for clarity) using $\approx\num{70}$ 
experimental realizations per temperature, plotted along with the corresponding 
equilibrium prediction (no free parameters).
    }
	\label{fig:modes}
\end{figure*}

\vspace{3pt}\noindent
\textit{Ripplon modes}---We begin with a data-driven analysis of the observed 
height profiles using principal component analysis 
(PCA)~\cite{Segal2010,Dubessy2014,Galka2022} to determine the mode structure 
from the statistically independent contributions of observed fluctuations.
This process constructs a basis of orthogonal components, indexed by an integer 
$n$ in order of decreasing variance (weight) across the dataset.
Figure~\ref{fig:modes}(a) shows example PCA components, with corresponding 
weights in Fig.~\ref{fig:modes}(b).
The PCA components closely resemble the cosinusoidal modes expected for 1D 
waves;
we computed the discrete cosine transform (DCT) of each component 
[Fig.~\ref{fig:modes}(c)] and found a nearly one-to-one correspondence between 
component index $n$ and dominant wavenumber $k$.
This supports the identification of interfacial fluctuations as occupied 
ripplon modes and underpins the DCT analysis outlined below.
At large wavenumber ($k / 2\pi > $ \qty{0.10}{\per\um}), imaging aberrations 
attenuated the observed amplitudes~\cite{ImagingFootnote}, reducing the 
signal-to-noise ratio of the PCA components and causing both imperfect 
correlation between PCA index $n$ and $k$ and mixing between $k$-components.

Interfacial modes such as ripplons are well described by the Hamiltonian
\begin{align}
H &= \frac{1}{L_0} \sum_k \frac{1}{2}\left[\frac{\hbar^2}{\rho_{\rm t}} |k| 
q^\dagger_k q_k + \sigma \left(\xi^{-2} +  k^2\right) h^\dagger_k 
h_k\right],\label{eq:quadratic_energy}
\end{align}
which expresses the dynamics of the Fourier transformed height field $h_k$ and 
its conjugate momentum $\hbar q_k$ in terms of the combined mass density of the 
two fluids $\rho_{\rm t}$, the interfacial tension $\sigma$, and the capillary 
length $\xi$ (all determined from solving the Gross-Pitaevskii and 
Bogoliubov-de Gennes equations~\cite{VanScha2008,GengTao2025}; see SM).
From the second term, the equipartition theorem predicts the 
thermal-equilibrium power spectral density
\begin{align}
     \frac{\langle |h_k|^2 \rangle}{L_0} &= \frac{\kB T}{\sigma (k^2 + 
\xi^{-2})}. \label{eq:pheno_c_k}
\end{align}
of the height field~\cite{Flekko1995,Aarts2004}; this prediction is a 
decreasing function of $k$, in qualitative agreement with the dependence of PCA 
weights in Fig.~\ref{fig:modes}(b,c).
To make the comparison quantitative, Fig.~\ref{fig:modes}(d) plots the measured 
ensemble averaged power spectral density $\langle |h_k|^2 \rangle$ (markers) 
along with the predictions of Eq.~\eqref{eq:pheno_c_k} (no adjustable 
parameters) for several different values of $T_{\rm bulk}$.
For all $T_{\rm bulk}$, the experimental $\langle |h_k|^2 \rangle$ follows a 
similar trend: high-$k$ points [plotted as filled markers] coincide with the 
equilibrium prediction, while low-$k$ points indicate a larger-than-expected 
population of ripplons.

\begin{figure}[t]
    \centering
    \includegraphics{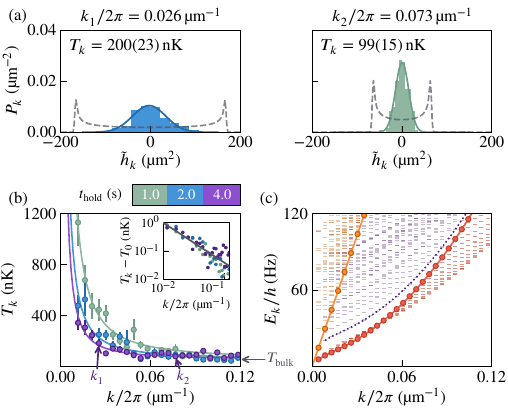}
    \caption{
    	Distribution functions and equilibration at $T_{\rm bulk} = 
\qty{56}{\nano\kelvin}$.
        (a) PDFs $P_k$ for $k_1/(2\pi) = \qty{0.026}{\per\um}$ and $k_2/(2\pi) 
= \qty{0.073}{\per\um}$ at $t_{\rm hold} = 4\ {\rm s}$.
        The histograms plot experimental data with corresponding normal 
distributions (solid curves).
        Dashed curves show the PDF corresponding to a phase-averaged ripplon 
coherent state with the same ripplon occupation.
        (b) Mode temperature as a function of $k$ for three hold times.
        Solid curves show power-law fits with a common exponent and offset.
        Inset: log-log plot in terms of dimensionless quantities $\bar{T}_k$ 
and $\bar{k}$ showing data collapse.
        (c) Scattering processes in our $L_0 \times L_0$ box.
        Circles connected by curves plot the energy of ripplons (red) and 
phonons (yellow) as a function $k$.
        Ticks indicate the total energy of momentum-conserving 3-field 
processes starting from a momentum-$k$ ripplon and ending in a pair of phonons 
(yellow), a phonon and a ripplon (purple), or two ripplons (red).
        The dotted curve traces out the lowest-energy phonon-producing channel 
for each $k$.
    } \label{fig:prethermal}
\end{figure}
\vspace{3pt}\noindent
\textit{Distribution functions}---To go beyond the average properties, we 
extract the full probability density function (PDF) $P_k \equiv P(h_k)$ of 
ripplon amplitudes from ensembles of about $70$ experimental realizations.
Figure~\ref{fig:prethermal}(a) shows representative low- and high-$k$ mode PDFs 
at $T_{\rm bulk} = \qty{56}{\nano\kelvin}$; the broader PDF at low-$k$ explains 
the $k$ dependence of $\langle |h_k|^2 \rangle$ observed in 
Fig.~\ref{fig:modes}(d) as they originate from the second-moment of these 
distributions.
Across the data set, $P_k$ is statistically indistinguishable from a normal 
distribution [solid curves in Fig.~\ref{fig:prethermal}(a)], with the moments 
$\langle h_k^m \rangle$ agreeing within expected uncertainties up to $m=6$ (for 
higher moments, the uncertainties exceed the mean values).
This Gaussian form is the expected thermal distribution for the quadratic 
ripplon Hamiltonian in Eq.~\eqref{eq:quadratic_energy}.
We therefore assign a temperature $T_k$ to each individual ripplon mode (this 
neglects quantum fluctuations: a valid approximation since $\hbar \omega_k \ll 
\kB T$, see SM).

Figure~\ref{fig:prethermal}(b) extends this analysis over the full range of $k$ 
and for hold times \qtyrange{1}{4}{\second}.
The resulting $T_k$ values show that only high-$k$ ripplon modes equilibrate 
with the bulk; at low-$k$, the mode temperatures exceed $T_{\rm bulk}$ and we 
empirically find that excessive temperature is well described by a power-law 
$(T_k - T_{\rm bulk})/T_0 = k^{-\alpha}$, with exponent $\alpha=1.42(5)$ and 
prethermal temperature scale $T_0$; the resulting data-collapse is shown in 
Fig.~\ref{fig:prethermal}(b).
These low-$k$ modes decay extremely slowly, with a \qty{5(1)}{\second} measured 
decay time---much slower than the $\sim \qty{e-1}{\second}$ equilibration 
timescale of the phonon bath, highlighting the prethermal nature of the 
interface system.
The excess low-$k$ occupation and slow relaxation might plausibly yield 
transient ripplon condensation~\cite{Fang2016}: where either a coherent state 
(averaged over an unknown phase) or a number state underlies the PDF.
For our large ripplon occupation, their resulting PDFs would be 
indistinguishable and are shown as dashed curves in 
Fig.~\ref{fig:prethermal}(a).
The possibility of ripplon condensation is excluded by the observed 
single-peaked Gaussian PDFs.

\vspace{3pt}\noindent
\textit{Mechanism}---As illustrated in Fig.~\ref{fig:schematics}, our 
prethermal state only appears after a well-defined interface develops, enabling 
a description in terms of ripplons.
Since this configuration is not far from the ground state, its excitations can 
be treated using the Bogoliubov-de Gennes (BdG) formalism~\cite{Pitaevskii2016} 
to identify weakly interacting quasi-particles: interfacial ripplons and bulk 
phonons.
The solid circles in Fig.~\ref{fig:prethermal}(c) plot the resulting energies 
for ripplons (red) along with linearly dispersing longitudinal phonons (orange).
We attribute a microscopic origin of the prethermal state to a decoupling of 
small $k$ ripplons from the phonons: the already weak interaction between BdG 
quasiparticles is further suppressed by the clear mismatch between their 
energies~\cite{Barnett2011,Gong2013,Yin2023}.
The lowest-order mechanisms by which ripplons couple to each other or decay 
into phonons are three-field processes---the scattering between a high-energy 
quasi-particle and a pair of low-energy quasi-particles or the reverse---that 
lead to Beliaev damping~\cite{Beliaev1957,Dereziski2024,Zhang2021}.

The tick marks in Fig.~\ref{fig:prethermal}(c)\ plot the total energy of 
quasiparticle pairs with total longitudinal momentum $k$: two phonons (yellow), 
a phonon and a ripplon (purple), and two ripplons (red).
Energy- and momentum-conserving transitions at this order (i.e. on-shell 
tree-level scattering) are only possible where ticks and circles coincide.
At low $k$ where excess ripplon occupation is observed, phonon-generating 
processes are gapped by $\approx 20\ {\rm Hz}$ (dotted curve), large compared 
to the ripplon sub-Hz linewidth.
This precludes these lowest-order on-shell scattering channels and provides a 
type of kinematic isolation of these modes from the phonon bath. At high $k$ 
this gap closes, potentially explaining why these ripplons rapidly equilibrate.

\begin{figure}[t]
    \centering
    \includegraphics{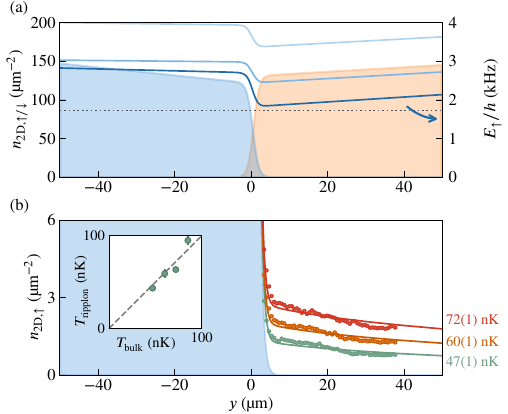}
    \caption{
    	Minority thermal atom thermometry.
        (a) Effective potentials for thermal spin-$\uparrow$ atoms in the 
lowest three local transverse modes (solid curves), compared with the chemical 
potential $\mu_\uparrow$ (dotted line).
        The shaded curves plot the condensate densities 
$n_{\uparrow/\downarrow}$ in blue/yellow respectively.
        (b) Spin-$\uparrow$ density profiles.
        Markers plot the measured total density (thermal plus condensed) along 
with fits (solid curve) to our theoretical model, yielding the annotated 
temperatures.
			The shaded region shows the computed condensate density 
and highlights its vast difference in scale.
        Inset: $T_{\rm bulk}$ versus $T_{\rm ripplon}$ obtained by fitting the 
high-$k$ ripplon tail in Fig.~\ref{fig:modes}(d) using Eq.~\eqref{eq:pheno_c_k} 
with $T_{\rm ripplon}$ as the only free parameter; the dashed line indicates 
equality.
    } \label{fig:thermal_density}
\end{figure}

\vspace{3pt}\noindent
\textit{Thermometry}---The above analysis requires an independent determination 
of the bulk temperature for the fit-parameter-free comparison shown in 
Fig.~\ref{fig:modes}(d).
Temperature measurement is a long-standing puzzle in quantum gas 
experiments~\cite{Fang2016, Ville2018}; in our context, the main difficulty is 
separating the minuscule contribution of thermal excitations from an enormous 
condensate background.
We address this issue by using spin-selective imaging to detect thermal 
excitations of one spin state that reside in the otherwise phase-separated 
condensate of the opposite spin state, yielding the faint box-shaped ``shadow'' 
visible in the upper half of Fig.~\ref{fig:schematics}(v, vi).

The density of thermal spin-$\uparrow$ atoms is given by $n_{T, \uparrow} = 
g_{1}[z(\bf{r})] / \lambda_T^2$, where: the fugacity $z(\mathbf{r}) 
=\exp[(V_{\rm eff}(\mathbf{r}) - \mu_\uparrow)/k_B T] $; $\mu_\uparrow$ is the 
spin-$\uparrow$ chemical potential; $\lambda_T
\equiv (2\pi\hbar^2/m k_B T)^{1/2}$ is the thermal de Broglie wavelength;
$g_{n}(z)$ are Bose functions~\cite{Giorgini1997}; and
\begin{align}
V_{\rm eff, \uparrow}(\mathbf r) &= V_{\rm ext}(\mathbf r) + 2 
g_{\uparrow\uparrow} n_{\uparrow}(\mathbf r) + g_{\uparrow\downarrow} 
n_{\downarrow}(\mathbf r), \label{eq:v_eff}
\end{align}
is an effective potential in terms of the external confinement and condensate 
densities $n_{\uparrow/\downarrow}$ [shaded regions in 
Fig.~\ref{fig:thermal_density}(a)].
The dark blue curve in Fig.~\ref{fig:thermal_density}(a) shows $V_{\rm eff}$ 
for the thermal minority atoms.
To obtain quantitatively accurate results in our quasi-2D configuration, we 
include the small occupation of excited transverse modes (see SM).
The factor of two in the same-spin interaction term is the Hartree-Fock 
exchange contribution and, at these low densities, self-interactions between 
thermal atoms are negligible.
Effective potentials for the lowest excited transverse modes are shown by the 
dark-blue curves in Fig.~\ref{fig:thermal_density}(a).

The expression for minority-spin density simplifies to $n_{T, \uparrow} = 
g_{1}[\exp(V_{\rm ext})/k_B T] / \lambda_T^2$ in the Thomas-Fermi limit where 
$\mu_{\uparrow} = g_{\uparrow\downarrow} n_{\downarrow}$; this removes 
sensitivity to the condensate density, making the minority thermal density an 
ideal thermometer.
Figure~\ref{fig:thermal_density}(b) presents the observed density of minority 
thermal atoms (colored circles) for three different temperatures; the 
comparison excludes the box edge where $V_{\rm eff,\uparrow}$ varies rapidly.
The fits (with only $T_{\rm bulk}$ as a free parameter) are in good agreement 
with observations.
The inset of Fig.~\ref{fig:thermal_density}(b) compares the $T_{\rm bulk}$ 
obtained via this procedure with $T_{\rm ripplon}$ extracted from fitting 
Eq.~\eqref{eq:pheno_c_k} to the high-$k$ tail of measured $\langle |h_k|^2 
\rangle$ [Fig.~\ref{fig:modes}(d)].
The good agreement between these two temperatures provides an independent 
validation of our thermometry technique.

\vspace{3pt}\noindent
\textit{Outlook}---We experimentally identified a prethermal subsystem of 
ripplon modes in a quenched binary BEC in which high-$k$ modes equilibrate with 
the condensate bulk, while low-$k$ modes remain described by effective 
temperatures greatly exceeding the bulk temperature.
These low-$k$ modes are weakly coupled to each other and bulk phonons through 
three- and four-field scattering processes, and only couplings to other low-$k$ 
ripplons satisfy energy and momentum conservation.
We therefore attribute the prethermal behavior to the lack of on-shell 
processes coupling low-$k$ ripplons to the phonon bath, analogous to the 
mechanism for slow thermalization in FPUT systems~\cite{Onorato2015}.

Although the case for prethermal behavior in this system is clear, many 
questions remain unanswered.
For example: what are the specific processes leading to the initial prethermal 
configuration, and what processes control its relaxation?  And why can 
restricted observables of individual ripplon mode amplitudes appear thermal 
even while the set of mode temperatures remains globally nonthermal?
Perhaps the answer is eigenstate thermalization hypothesis 
adjacent~\cite{Rigol2008,DAlessio2016}; alternatively, the thermal single-mode 
statistics may reflect dephasing or kinetic mixing within the prethermal 
subsystem, closer to generalized-ensemble descriptions of restricted 
observables~\cite{Gring2012,Langen2015,Mori2018}.
Overall, we believe our experimental platform opens a route toward a more 
complete understanding of prethermalization across a broader class of quantum 
systems.

\begin{acknowledgments}
The authors thank S.~Mukherjee and Q. Yao for technical assistance, and  
V.~Galitski for productive discussions.
This work was partially supported by the National Institute of Standards and 
Technology; the National Science Foundation through the Quantum Leap Challenge 
Institute for Robust Quantum Simulation (grant OMA-2120757); and the Air Force 
Office of Scientific Research Multidisciplinary University Research Initiative 
``RAPSYDY in Q'' (No. FA9550-22-1-0339).
\end{acknowledgments}
\bibliography{arxiv}

\newpage
\onecolumngrid

\section{Supplemental Material}
\def\PsiOp{\hat\Psi(\mathbf{r})}
\def\PsiOps#1{\hat\Psi_{#1}(\mathbf{r})}
\def\PsiOpT{\hat\Psi(\mathbf{r}, t)}
\def\PsiDagOp{\hat\Psi^\dagger(\mathbf{r})}
\def\PsiDagOps#1{\hat\Psi_{#1}^\dagger(\mathbf{r})}
\def\PsiDagOpT{\hat\Psi^\dagger(\mathbf{r}, t)}
\def\vecr{\mathbf{r}}
\def\vecp{\mathbf{p}}
\def\xvec{\hat{\mathbf{x}}}
\def\yvec{\hat{\mathbf{y}}}
\def\zvec{\hat{\mathbf{z}}}
\def\dvol{\mathrm d^3 \mathbf{r}}
\def\upup{{\uparrow\uparrow}}
\def\downdown{{\downarrow\downarrow}}
\def\updown{{\uparrow\downarrow}}
\def\aOp{\hat a}
\def\aOpDag{\hat a^\dagger}
\def\bOp{\hat b}
\def\bOpDag{\hat b^\dagger}
\def\kB{k_\mathrm{B}}

\appendix

\setcounter{equation}{0}
\setcounter{figure}{0}
\setcounter{table}{0}

\renewcommand\theequation{S\arabic{equation}}
\renewcommand\thefigure{S\arabic{figure}}

\subsection{Normal-mode relations for the interface}

The capillary-wave model in the main text uses the Fourier amplitude $h_k$ of 
the interface height field.
We fix the normalization of $h_k$ and motivate the effective interface model by 
first deriving the normal modes of a 1D string with linear mass density $\rho$ 
and tension $\sigma$.

\noindent
\textit{Basic model}---We model the interface as a chain of $N$ oscillators in 
a 1D ring undergoing only vertical motion.
Interfacial tension, with strength $\sigma$, and buoyancy, with differential 
force per area $\Delta f$, give the potential energy:
\begin{align*}
V &= \sum_j \left[\sigma \sqrt{\delta x^2 + (h_j - h_{j+1})^2}  + 
\frac{1}{2}\Delta f \delta x h_j^2\right]\approx \sum_j\left[ \sigma\delta x + 
\frac{\sigma}{2 \delta x} (h_j - h_{j+1})^2 + \frac{1}{2}\Delta f \delta x 
h_j^2\right] \\
&\rightarrow \sum_j\frac{1}{2} \left[m \omega^2(h_j - h_{j+1})^2 + \Delta f 
\delta x h_j^2 \right],
\end{align*}
Here $h_j$ is the height at site $j$; $\delta x$ is the spacing; $m = \rho 
\delta x$ is the oscillator mass; and $\omega = \sqrt{\kappa/m}$ is the 
oscillator frequency for a spring constant $\kappa = \sigma/\delta x$.
We drop the overall energy offset from the $\propto \sigma \delta x$ term.

The corresponding Lagrangian and Hamiltonian are:
\begin{align*}
L &= \sum_j \frac{1}{2}\left[ m v_j^2 - m \omega^2 (h_j - h_{j+1})^2 - \Delta f 
\delta x h_j^2\right], & \mathrm{and} &&
H &= \sum_j \frac{1}{2}\left[\frac{\hbar^2}{m} q_j^2 + m \omega^2 (h_j - 
h_{j+1})^2 + \Delta f \delta x h_j^2\right],
\end{align*}
Here $v_j$ is the velocity, and $\hbar q_j$ is the conjugate momentum of $h_j$.
We then quantize $h_j$ and $q_j$ with $[h_{j},q_{j'}]=i \delta_{j,j'}$ and use 
the Fourier-transform convention:
\begin{align*}
h_k &= \frac{1}{\sqrt{N}} \sum_j e^{ikj} h_j & {\rm and} && h_j &= 
\frac{1}{\sqrt{N}} \sum_k e^{-ikj} h_k,
\end{align*}
The same convention applies to $q_k$.
These variables are not Hermitian, and they have the commutation relation 
$[h_{k},q^\dagger_{k'}]=[h_{k},q_{-k'}]=i \delta_{k,k'}$.

In terms of these variables, the Hamiltonian is:
\begin{align*}
H &= \sum_k \frac{1}{2}\left[\frac{\hbar^2}{m} q^\dagger_k q_k + m \omega_k^2 
h^\dagger_k h_k\right],
\end{align*}
This form exposes the normal-mode structure, with $\omega_k^2 = 4 \omega^2 
\sin^2\left(k/2\right) + \Delta f \delta x / m$.
The corresponding bosonic annihilation operator is:
\begin{align*}
a_k &= \frac{1}{\sqrt{2}}\left(\sqrt{\frac{m \omega_k}{\hbar}} h_k + 
i\sqrt{\frac{\hbar}{m \omega_k}} q^\dagger_k\right)
\end{align*}
The canonical variables are then:
\begin{align*}
h_k &= \sqrt{\frac{\hbar}{m \omega_k}} \frac{a_k + a^\dagger_{-k}}{\sqrt{2}} & 
\mathrm{and} && q^\dagger_k &= \sqrt{\frac{m \omega_k}{\hbar}} \frac{a_k- 
a^\dagger_{-k}}{\sqrt{2} i},
\end{align*}
The Hamiltonian is $H = \sum_k \hbar \omega_k (a^\dagger_k a_k + 1/2)$.
This Fourier-transform convention leaves the dimensions unchanged, so $h_k$ 
still has dimensions of $[\mathrm L]$.

\noindent
\textit{Continuum limit}---The continuum limit requires more care.
For a system with length $L_0 = N \delta x$, the Lagrangian becomes the Riemann 
sum:
\begin{align*}
L &= \sum_j \delta x \frac{1}{2} \left[ \rho v_j^2 - \sigma \left(\frac{h_j - 
h_{j+1}}{\delta x}\right)^2 -  \Delta f h_j^2\right] \rightarrow \int dx 
\frac{1}{2}\left\{\rho v(x)^2 - \sigma \left[\partial_x h(x)\right]^2-\Delta f 
h(x)^2\right\},
\end{align*}
The kinetic, elastic, and restoring terms are parameterized by intensive 
quantities: the 1D density $\rho$, the interfacial tension $\sigma$, and the 
differential force per area $\Delta f$.
We started with the Lagrangian because it identifies the conjugate variable of 
$h(x)$ via $\hbar q(x) = \delta L / \delta v(x) = \rho v(x)$; here $\hbar q(x)$ 
is the momentum density.
The Hamiltonian is therefore:
\begin{align*}
H &\rightarrow \int dx \frac{1}{2}\left\{\frac{\hbar^2}{\rho}q(x)^2 + \sigma 
\left[\partial_x h(x)\right]^2 + \Delta f h(x)^2\right\},
\end{align*}
It obeys $[h(x), q(x')]=i \delta(x - x')$.
Dimensional analysis gives $[h(x)] = [\mathrm{L}]$ and $[q(x')] = 
[\mathrm{L}]^{-2}$.
We use the Fourier transform pair:
\begin{align*}
h_k &= \int_0^{L_0} dx e^{ikx} h(x) & {\rm and} && h(x) &= \frac{1}{L_0} \sum_k 
e^{-ikx} h_k,
\end{align*}
The $k$-sum goes from $-\infty$ to $+\infty$ with steps of $\delta k = 2\pi / 
L_0$, and yields the commutation relation 
$[h_{k},q^\dagger_{k'}]=[h_{k},q_{-k'}]=i L_0 \delta_{k,k'}$.
This convention converges to the Dirac delta function as the system size 
diverges because the expression for $h(x)$ is the Riemann sum leading to $h(x) 
= \int dk / (2\pi) \exp(-ikx) h(k)$.
Now $h_k$ has dimensions of $[\mathrm L]^{2}$ and $q_k$ has dimensions of 
$[\mathrm L]^{-1}$.

The Hamiltonian in Fourier variables is:
\begin{align*}
H &= \frac{1}{L_0} \sum_k \frac{1}{2}\left[\frac{\hbar^2}{\rho} q^\dagger_k q_k 
+ (\sigma k^2 + \Delta f) h^\dagger_k h_k \right] = \frac{1}{L_0} \sum_k 
\frac{1}{2}\left(\frac{\hbar^2}{\rho} q^\dagger_k q_k + \rho \omega_k^2 
h^\dagger_k h_k\right),
\end{align*}
The mode frequency is $\omega^2_k = (k^2 \sigma + \Delta f)/\rho$.
We now seek bosonic operators with the commutation relation $[a_k, 
a^\dagger_{k'}] = \delta_{k,k'}$ yielding $H = \sum_k \hbar \omega_k 
(a^\dagger_k a_k + 1/2)$.

With this convention, the canonical variables are:
\begin{align*}
h_k &= L_0 \sqrt{\frac{\hbar}{L_0 \rho \omega_k}} \frac{a_k + 
a^\dagger_{-k}}{\sqrt{2}} & \mathrm{and} && q^\dagger_k &= \sqrt{\frac{L_0 \rho 
\omega_k}{\hbar}} \frac{a_k- a^\dagger_{-k}}{\sqrt{2} i}.
\end{align*}

\textit{Connection to interface problem}---The height part of the Hamiltonian 
in the main text is:
\begin{align}
    H_h &= \frac{1}{L_0} \sum_k \frac{1}{2} \sigma (k^2 + \xi^{-2}) h^\dagger_k 
h_k, \label{eq:height_part}
\end{align}
Here $\xi^2 = \sigma/\Delta f$ is the capallary length~\cite{Aarts2004}, for 
differential force per unit area $\Delta f$.
In our system, $\Delta f = \mu_{\rm B} (g_\uparrow n_\uparrow - g_\downarrow 
n_\downarrow) B'_z$ is the areal force density (analogous to the volumetric 
force density in a 3D fluid), where $\mu_{\rm B}$ is the Bohr magneton; and 
$n_{\uparrow,\downarrow}$ and  $g_{\uparrow,\downarrow}$ are spin-dependent 
condensate density and Land\'{e} $g$-factors.
The capillary length is $1/\xi = 2\pi \times \qty{0.038(1)}{\per\um}$ in our 
experiment.
This model assumes a single-valued $\sigma$ for all $k$, which is approximately 
true for a wide range of $k$ in our system.
For a local string this would imply the mode frequency $\omega^2_k = 
\left(\sigma / \rho_k\right)\times(k^2 + \xi^{-2})$.
This local-string result lacks the nonlocal inertia of an interface wave and 
therefore does not match the known dispersion $\omega^2_k = (\sigma/\rho_{\rm 
t}) |k| \left(\xi^{-2} +  k^2\right)$, where $\rho_{\rm t} = \rho_\uparrow + 
\rho_\downarrow$ is the total 2D mass density.

To reproduce this dispersion, we use the mode-dependent density:
\begin{align*}
\rho_k &= \frac{\rho_{\rm t}}{|k|}.
\end{align*}
This scaling reflects the boundary layer of fluid, with width of order 
$k^{-1}$, that co-moves with the wave.
A box of finite height $D$ introduces a cutoff, so for $k D \lesssim 1$, 
$\rho_k$ saturates around $\rho_k\simeq\rho_{\rm t} D$.
This identifies the quantity $\eta_k = \sqrt{\hbar / (L_0\rho_k \omega_k)}$ 
defined above.

With this mode-dependent inertia, we arrive at the total Hamiltonian 
[Eq.~\eqref{eq:quadratic_energy}]:
\begin{align*}
H &= \frac{1}{L_0} \sum_k \frac{1}{2}\left[\frac{\hbar^2}{\rho_{\rm t}} |k| 
q^\dagger_k q_k + \sigma \left(\xi^{-2} +  k^2\right) h^\dagger_k h_k\right].
\end{align*}

To arrive at Eq.~\eqref{eq:pheno_c_k}, we focus on the height part of the 
Hamiltonian and invoke the equipartition theorem by setting the expectation of 
Eq.~\eqref{eq:height_part} to $k_B T / 2$~\cite{Flekko1995,Aarts2004}. It is 
worth noting that, other than being quadratic in $q_k$, the exact form of the 
momentum part in Eq.~\eqref{eq:quadratic_energy}  is not necessary to derive 
Eq.~\eqref{eq:pheno_c_k}.

\subsection{Quantum theory of ripplons in a binary condensate}

The weakly interacting binary Bose gas is described by the Hamiltonian:
\begin{align*} \hat H = \int {\mathrm d}^3 {\mathbf r}
    &\left\{  \hat \Psi_{\uparrow}^{\dagger}({\mathbf r}) \left[ -\frac{\hbar
            ^{2}}{2 m} \nabla^2 +V_\uparrow\left( {\mathbf r}\right) \right]
            \hat \Psi_{\uparrow}({\mathbf r}) + \hat \Psi
            _\downarrow^{\dagger}({\mathbf r}) \left[ -\frac{\hbar ^{2}}{2 m}
        \nabla^2+V_\downarrow\left( {\mathbf r}\right) \right] \hat
    \Psi_{\downarrow}({\mathbf r}) \right.\\
    & +\frac{g_{\upup}}{2}\Psi _{\uparrow}^{\dagger}({\mathbf r})\Psi
    _{\uparrow}^{\dagger}({\mathbf r}) \hat \Psi_{\uparrow}({\mathbf r}) \hat
    \Psi_{\uparrow}({\mathbf r}) +\frac{g_{\downdown}}{2}\Psi
    _{\downarrow}^{\dagger}({\mathbf r})\Psi _{\downarrow}^{\dagger}({\mathbf
    r}) \hat \Psi_{\downarrow}({\mathbf r}) \hat \Psi_{\downarrow}({\mathbf r})
    \\
    &+ \left. g_{\updown}\Psi _{\uparrow}^{\dagger}({\mathbf r})\Psi
_{\downarrow}^{\dagger}({\mathbf r}) \hat \Psi_{\downarrow}({\mathbf r}) \hat
\Psi_{\uparrow}({\mathbf r}) \right\}.
\end{align*}
Here $\hat \Psi_{\uparrow,\downarrow}$ are the bosonic field operators, 
$g_\upup, g_\downdown$ are the intra-spin interaction strengths, and 
$g_\updown$ is the inter-spin interaction strength~\cite{Lamporesi2023}.
We compute the ground state of the two-dimensional (2D) binary Bose-Einstein 
condensate (BEC) by replacing $\PsiOps{\uparrow,\downarrow} \to 
\psi_{\uparrow,\downarrow}(\vecr)$ and minimizing the system energy over 
$\psi_{\uparrow,\downarrow}(\vecr)$.
The resulting $\psi_{0\uparrow,\downarrow}(\vecr)$ is the condensate 
ground-state wavefunction.

We obtain the quasiparticle excitation spectrum by expanding the Hamiltonian to 
second order in small perturbations around the ground state: 
$\PsiOps{\uparrow,\downarrow} = \psi_{0\uparrow,\downarrow}(\vecr) + \delta 
\PsiOps{\uparrow,\downarrow}$.
The resulting Hamiltonian is diagonalized by the Bogoliubov 
transformation~\cite{Pitaevskii2016}:
\begin{align*}
    \delta \PsiOps{\uparrow/\downarrow}
    &= \sum_i \left[ u_{\uparrow/\downarrow, i}(\vecr) \bOp_i - 
v_{\uparrow/\downarrow, i}^*(\vecr) \bOpDag_i \right], \\
    \delta \PsiDagOps{\uparrow/\downarrow}
    &= \sum_i \left[ u_{\uparrow/\downarrow, i}^*(\vecr) \bOpDag_i - 
v_{\uparrow/\downarrow, i}(\vecr) \bOp_i \right],
\end{align*}
Here $\hat b$ is the quasiparticle operator.

For each momentum $k$, we numerically solve for $u_{\uparrow/\downarrow, 
i}(\vecr)$ and $v_{\uparrow/\downarrow, i}(\vecr)$ using the ansatz below.
The ansatz is $u_{\uparrow/\downarrow, i}(x, y) = u_{\uparrow/\downarrow, i}(y) 
\exp(ikx) / \sqrt{L_0}$ and $v_{\uparrow/\downarrow, i}(x, y) = 
v_{\uparrow/\downarrow, i}(y) \exp(ikx)/ \sqrt{L_0}$.
We identify ripplons as the lowest-energy quasiparticles whose $u, v$ 
amplitudes are localized at the interface [Fig.~\ref{fig:prethermal}(c)].
We denote $\bOp_{0,k}$ and $\bOpDag_{0,k}$ as the operators that annihilate and 
create a ripplon at momentum $k$, with energy $\hbar \omega_k$.
The interface height operator can be constructed as $\hat h(x) = \sum_k \eta_k 
[\bOp_{0,k} \exp(ikx) + \bOpDag_{0,k}\exp(-ikx)]$, and its Fourier transform is 
$\hat h_k = L_0 \eta_k (\bOpDag_{0,k} + \bOp_{0,-k})$.
We compute $\eta_k$ from the shift of the interface per ripplon added to the 
ground-state wavefunction, yielding power spectral density $\langle |\hat 
h_k|^2 \rangle = \eta_k^2 L_0^2 (n_{0,k} + n_{0,-k} + 1)$, where $n_{0,k}$ is 
the ripplon occupation number at $k$.

\begin{figure}[t]
    \centering
    \includegraphics{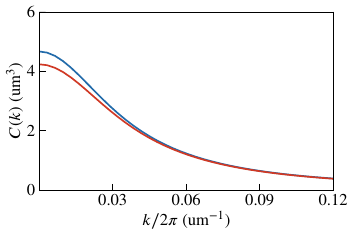}
    \caption{ Classical and quantum predictions for $C(k)$. Red and blue curves 
show the classical prediction [Eq.~\eqref{eq:pheno_c_k}] and quantum prediction 
[Eq.~\eqref{eq:bdg_c_k}] at \qty{80}{\nano\kelvin}, a typical experimental 
temperature. }
    \label{fig:c_k_theory}
\end{figure}

At thermal equilibrium, the ripplon occupation follows the Planck distribution 
$n_{0,k} = 1/[\exp(\hbar \omega_k/\kB T) - 1]$, where $\kB$ is the Boltzmann 
constant, and $T$ is the system temperature.
The resulting power spectral density is:
\begin{equation}
    \langle |\hat h_k|^2 \rangle = \eta_k^2 L_0^2 \coth \left( \frac{\hbar 
\omega_k}{2 \kB T} \right).
    \label{eq:bdg_c_k}
\end{equation}
Figure~\ref{fig:c_k_theory} compares $C(k)$ from Eq.~\eqref{eq:pheno_c_k} and 
Eq.~\eqref{eq:bdg_c_k}.

\subsection{Probability distribution of \texorpdfstring{$h_k$}{hk} for a 
quantum interface at thermal equilibrium}

Classically, the quadratic energy in Eq.~\eqref{eq:quadratic_energy} gives a 
normal distribution of $h_k$ for each ripplon mode in thermal equilibrium.
\begin{align*}
    P(h_k) = \exp\left[-\frac{\sigma (k^2 + \xi^{-2}) |h_k|^2}{2 L_0 \kB 
T}\right].
\end{align*}

\begin{figure}[b]
    \centering
    \includegraphics{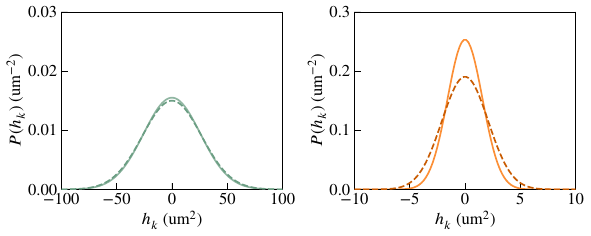}
    \caption{ Classical and quantum probability distributions for $h_k$. Solid 
and dashed curves show the classical prediction and 
Eq.~\eqref{eq:h_prob_quantum}, respectively, at \qty{80}{\nano\kelvin} (green, 
typical experiment temperature) and \qty{0.3}{\nano\kelvin} (orange). }
\label{fig:h_prob_compare}
\end{figure}

For the quantum case, the lowest-order ripplon Hamiltonian is $\hat H = \hbar 
\sum_k \omega_k \bOpDag_{0,k} \bOp_{0,k}$.
For the ripplon mode at a fixed $k$, the thermal density matrix is $\hat \rho_k 
= (1/Z) \sum_m \exp(-m \hbar \omega_k / \kB T) \ket{m} \bra{m}$.
The corresponding probability distribution of $h_k$ is given by $P(h_k) = 
\bra{h_k}\hat \rho_k\ket{h_k}$, where $\ket{m}$ is the number basis of the mode 
at $k$ and $\ket{h_k}$ is its ``displacement" basis, analogous to a quantum 
harmonic oscillator.
The probability distribution is therefore:
\begin{align*}
    P(h_k)
    &= \frac{1}{Z} \sum_m e^{-m \hbar \omega_k / \kB T}
    |\langle h_k | m \rangle|^2 \\
    &= \frac{1}{Z} \sum_m e^{-m \hbar \omega_k / \kB T}
    \frac{1}{2^m m!} \frac{1}{\sqrt{2\pi} \eta_k L_0}
    \exp \left(-\frac{|h_k|^2}{2 \eta_k^2 L_0^2}\right) H_m^2\left( 
\frac{h_k}{\sqrt{2}
    \eta_k L_0} \right),
\end{align*}
Using Mehler's 
formula~\cite[\href{http://dlmf.nist.gov/18.18.E28}{(18.18.28)}]{DLMF}, this 
becomes:
\begin{equation}
\begin{aligned}
    P(h_k)
    &\propto \exp \left[
        -\tanh\left(\frac{\hbar \omega_k}{2 \kB T}\right)
        \frac{1}{2 \eta_k^2 L_0^2} |h_k|^2
    \right], \\
    &= \exp \left( -\frac{\hbar \omega_k}{4 \kB T \eta_k^2 L_0^2} |h_k|^2 
\right)
    \quad \text{as}~\frac{\hbar \omega_k}{\kB T} \to 0.
\end{aligned}
\label{eq:h_prob_quantum}
\end{equation}
Thus the quantum distribution is also Gaussian.
The second moment of this probability distribution gives Eq.~\eqref{eq:bdg_c_k} 
and reduces to the classical equipartition result in the high-temperature limit 
$\hbar \omega_k / \kB T \to 0$.
Figure~\ref{fig:h_prob_compare} compares the classical equipartition prediction 
with Eq.~\eqref{eq:h_prob_quantum}.

\subsection{Calculating \texorpdfstring{$n_T$}{nT} in the quasi-2D potential}

\begin{figure}[t]
    \centering
    \includegraphics{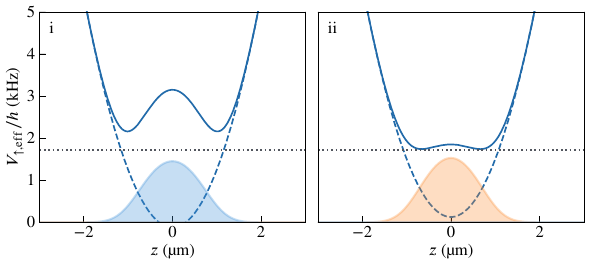}
    \caption{ Effective-potential slices used to calculate $n_T$. Left and 
right panels show $V_\text{eff}$ along $\mathbf e_z$ at $y = \qty{-31}{\um}$ 
(same-spin condensate region) and $y = \qty{31}{\um}$ (opposite-spin condensate 
region).}
    \label{fig:v_eff_z}
\end{figure}

The thermal atom density $n_T$ can often be computed with a simplified quasi-2D 
treatment when $\kB T \ll \hbar \omega_z$~\cite{Giorgini1997}.
In our experiment, however, $\hbar \omega_z$ and $\kB T$ are of the same order 
of magnitude.
A full quantum treatment of the transverse motion is therefore required.
We assume the following two conditions.
(1) $n_T$ is low enough so that the interaction between thermal atoms is 
negligible;
(2) $V_{\rm eff}$ varies slowly enough in $\mathbf e_y$, an approximation that 
fails at the edge of the box potential.
These two conditions reduce the problem to a 1D single-particle problem 
governed by the Schr\"{o}dinger equation (the $\partial^2_y$ term is dropped) 
in the potentials shown in Fig.~\ref{fig:v_eff_z}.

We numerically solve for the single-particle wavefunction $\phi_n(z)$ and 
energy $\epsilon_n$ in these potentials, then sum the probability density 
weighted by the Bose distribution to obtain:
\begin{equation}
\begin{aligned}
    n_T(\mathbf r)
    &= \sum_{n} |\phi_n(z)|^2 \int \frac{\mathrm d^2 \mathbf
    p}{(2\pi\hbar)^2} \left\{\exp\left[\frac{p^2/(2m) + \epsilon_n - \mu}{\kB 
T}\right] - 1\right\}^{-1}
    \\
    &= \sum_{n} |\phi_n(z)|^2 \frac{1}{\lambda_T^2}
    g_1\left[e^{(\mu - \epsilon_n)/\kB T}\right]
\end{aligned}
\label{eq:n_T}
\end{equation}
Here $g_n(z)$ are Bose functions and $\lambda_T \equiv [2\pi\hbar^2/(m \kB 
T)]^{1/2}$ is the thermal de Broglie wavelength.
We use $n_T$ from Eq.~\eqref{eq:n_T} as the fit model for the bulk temperature, 
producing the solid curves in Fig.~\ref{fig:thermal_density}(b).

\end{document}